\documentclass[12pt,preprint]{aastex}

\shorttitle{Scattering Polarization in the Hydrogen Lyman-$\alpha$ Line}
\shortauthors{Kano et al.}

\begin{document}

\title{Discovery of Scattering Polarization in the Hydrogen Lyman-$\alpha$ Line of the Solar Disk Radiation}

\author{R. Kano\altaffilmark{1,20}, 
	J. Trujillo Bueno\altaffilmark{2,3,4}, 
	A. Winebarger\altaffilmark{5}, 
	F. Auch\`{e}re\altaffilmark{6}, 
	N. Narukage\altaffilmark{1}, 
	R. Ishikawa\altaffilmark{1}, 
	K. Kobayashi\altaffilmark{5}, 
	T. Bando\altaffilmark{1}, 
	Y. Katsukawa\altaffilmark{1}, 
	M. Kubo\altaffilmark{1}, 
	S. Ishikawa\altaffilmark{7}, 
	G. Giono\altaffilmark{1,8}, 
	H. Hara\altaffilmark{1}, 
	Y. Suematsu\altaffilmark{1},
	T. Shimizu\altaffilmark{7}, 
	T. Sakao\altaffilmark{7}, 
	S. Tsuneta\altaffilmark{7}, 
	K. Ichimoto\altaffilmark{9}, 
	M. Goto\altaffilmark{10}, 
	L. Belluzzi\altaffilmark{11,12}, 
	J. \v{S}t\v{e}p\'{a}n\altaffilmark{13}, 
	A. Asensio Ramos\altaffilmark{3}, 
	R. Manso Sainz\altaffilmark{14},
	P. Champey\altaffilmark{15},
	J. Cirtain\altaffilmark{16}, 
	B. De Pontieu\altaffilmark{17}, 
	R. Casini\altaffilmark{18}, and 
	M. Carlsson\altaffilmark{19}}

\altaffiltext{1}{National Astronomical Observatory of Japan, 
	2-21-1 Osawa, Mitaka, Tokyo 181-8588, Japan}
\altaffiltext{2}{Instituto de Astrof\'{i}sica de Canarias, 
	La Laguna, Tenerife, E-38205, Spain} 
\altaffiltext{3}{Departamento de Astrof\'isica, Universidad de La Laguna, E-38206 La Laguna, Tenerife, Spain}
\altaffiltext{4}{Consejo Superior de Investigaciones Cient\'\i ficas, Spain} 	
\altaffiltext{5}{Marshall Space Flight Center, National Aeronautics 
	and Space Administration (NASA), Huntsville, AL 35812, USA}
\altaffiltext{6}{Institut d'Astrophysique Spatiale, 
	Universit\'{e} Paris Sud, Batiment 121, F-91405 Orsay, France}
\altaffiltext{7}{Institute of Space and Astronautical Science, 
	Japan Aerospace Exploration Agency, 
	3-1-1 Yoshinodai, Chuo, Sagamihara, Kanagawa 252-5210, Japan}
\altaffiltext{8}{The Graduate University for Advanced Studies (Sokendai), 
	Hayama, Kanagawa 240-0193, Japan}
\altaffiltext{9}{Hida Observatory, Kyoto University, 
	Takayama, Gifu 506-1314, Japan}
\altaffiltext{10}{National Institute for Fusion Science, 
	Toki, Gifu 509-5292, Japan}
\altaffiltext{11}{Istituto Ricerche Solari Locarno, 
	CH-6605 Locarno Monti, Switzerland}
\altaffiltext{12}{Kiepenheuer-Institut f\"ur Sonnenphysik, D-79104 Freiburg, Germany}
\altaffiltext{13}{Astronomical Institute, Academy of Sciences of 
	the Czech Republic, 25165 Ondrejov, Czech Republic}
\altaffiltext{14}{Max-Planck-Institut f\"ur Sonnensystemforschung, Justus-von-Liebig-Weg 3, D-37077 G\"ottingen, Germany}
\altaffiltext{15}{University of Alabama in Huntsville, 301 Sparkman Drive, Huntsville, AL 35899, USA}
\altaffiltext{16}{University of Virginia, Department of Astronomy, 530 McCormick Road, Charlottesville, VA 22904, USA}
\altaffiltext{17}{Lockheed Martin Solar \& Astrophysics Laboratory, 
	Palo Alto, CA 94304, USA}
\altaffiltext{18}{High Altitude Observatory, National Center for 
	Atmospheric Research, Post Office Box 3000, Boulder, CO 80307-3000, USA}
\altaffiltext{19}{University of Oslo, Postboks 1029 Blindern, NO-0315 Oslo, Norway}
\altaffiltext{20}{E-mail: ryouhei.kano@nao.ac.jp}

\begin{abstract}
There is a thin transition region (TR) in the solar atmosphere where the temperature rises 
from 10,000 K in the chromosphere to millions of degrees in the corona. Little is known about 
the mechanisms that dominate this enigmatic region other than the magnetic field plays a key role.
The magnetism of the TR can only be detected by polarimetric measurements of a few ultraviolet (UV) spectral lines, 
the Lyman-$\alpha$ line of neutral hydrogen at 121.6~nm (the strongest line of the solar UV spectrum) being of 
particular interest given its sensitivity to the Hanle effect (the magnetic-field-induced 
modification of the scattering line polarization). We report the discovery of 
linear polarization produced by scattering processes in the Lyman-$\alpha$ line, 
obtained with the Chromospheric Lyman-Alpha Spectro-Polarimeter (CLASP) rocket experiment. 
The Stokes profiles observed by CLASP in quiet regions of the solar disk show that the $Q/I$ and $U/I$ 
linear polarization signals are of the order of 0.1 \% in the line core and up to a few percent in the nearby wings, 
and that both have conspicuous spatial variations with scales of $\sim 10$~arcsec. 
These observations help constrain theoretical models of the chromosphere--corona TR and extrapolations 
of the magnetic field from photospheric magnetograms. In fact, the observed spatial variation from disk to limb of polarization at the line core and wings already challenge the predictions from three-dimensional magnetohydrodynamical models of the upper solar chromosphere. 
\end{abstract}


\section{Introduction}

The upper layers of the solar chromosphere, and the transition region (TR) in particular, are 
key to understanding several of the most fundamental problems in astrophysics such as the source 
of the solar wind, the energization of the corona, or the acceleration of energetic particles \citep[e.g.,][]{priest2014}.
Most of what we know about this puzzling atmospheric region has been learned from ground-based observations 
in the H$\alpha$ line \citep[e.g.,][]{kneer2010,rutten2017} and by combining space-based observations of the radiation 
intensity in ultraviolet (UV) spectral lines and magnetohydrodynamical numerical simulations 
\citep[e.g.,][]{hansteen2014,pereira2014}. Yet, we lack quantitative information on the single most 
important physical parameter: its magnetic field. To explore the magnetic field of the upper solar chromosphere and TR 
requires spectropolarimetry, i.e., the ability to detect the polarization state of photons  
in magnetically sensitive spectral lines, at UV wavelengths where photons are relatively 
scarce. In particular, the magnetic field of the rarified plasma 
of the chromosphere--corona TR can be probed only through a handful of UV spectral lines sensitive 
to the Hanle effect \citep[the magnetic-field-induced modification of the line scattering polarization; see][]{LL04},
but their observation and modeling are not easy \citep[e.g.,][]{trujillo2014}. 
On the one hand, these wavelengths cannot be observed 
from the ground, which poses a clear instrumental and observational challenge; on the other, the 
formation of the polarization profiles of these lines involves complex atomic and radiative 
mechanisms whose impact is only known theoretically, since the physical conditions of the TR are not 
accessible in terrestrial laboratories. These difficulties 
have left the strength and geometry of the magnetic field in 
the numerical models of the upper solar chromosphere largely unconstrained. Unable 
to directly confront models to observable quantities sensitive to the magnetic field, complex 
extrapolations of solar surface magnetograms are instead used, which are problematic because 
they are difficult or impossible to validate against actual observations \citep[e.g.,][]{derosa2009}. 

The dominant spectral feature in the UV solar spectrum is the hydrogen Lyman-$\alpha$ emission 
line at 121.6 nm, which forms all through the upper chromosphere and in the TR. This resonance line is 
dominated by scattering, and hence it could show a linear polarization pattern sensitive to the 
geometry of the formation region and to the presence of magnetic fields through the Hanle 
effect. A first---and for almost 40 years, only---attempt to detect solar limb polarization 
in this line using a slit-less spectropolarimeter failed, 
presumably due to in-flight degradation of its efficiency \citep{stenflo1980}. 
Recent theoretical work has predicted that in models of the solar atmosphere, with 
magnetic fields between 10 and 100~Gauss, the measurable level of linear polarization in the 
Lyman-$\alpha$ radiation of the solar disk should be of the order of 0.1~\% in the line core and up to a 
few percent in the nearby wings \citep{trujillobueno2011,belluzzi2012,stepan2015}.

In order to achieve the first measurement of the linear polarization produced by scattering 
processes in the hydrogen Lyman-$\alpha$ line radiation of the solar disk, we have developed 
the Chromospheric Lyman-Alpha Spectro-Polarimeter \citep[CLASP;][]{kano2012,kobayashi2012}. 
CLASP was successfully launched from White Sands Missile Range by a NASA sounding 
rocket on 2015 September 3, and the aim of this Letter is to show and discuss the results of these unprecedented 
spectropolarimetric observations.

\section{The instrument}

The CLASP 
\citep{narukage2015,narukage2017} consists of a Cassegrain telescope with an aperture of 27 cm; a spectropolarimeter to measure the wavelength variation of the Stokes $I$, $Q/I$ and $U/I$ profiles along the 400~arcsec covered by 
the spectrograph's slit (nearly half the solar radius); and a slit-jaw optical system to take 
fast-cadence (0.6 sec) chromospheric images with a Lyman-$\alpha$ broadband filter (FWHM=7 nm). 
The spectropolarimeter uses a rotating half-waveplate for 
the Lyman-$\alpha$ wavelength \citep{ishikawa2013,ishikawa2015}, located in front of the slit, 
which modulates the polarization of the incident radiation 4 times per rotation (one rotation takes 4.8 s). 
It also uses a single concave diffraction grating, mounted in the inverse Wadsworth configuration, 
which serves both as the spectral dispersion element and beam splitter \citep{narukage2015}. 
Each of the resulting optically symmetric channels goes through a 
linear-polarization analyzer mounted $90^{\circ}$ from each other to simultaneously measure 
two orthogonal polarization states. From the observed signals in both channels, we derive 
the wavelength variation of the Stokes $I$ profile and of the fractional polarization ($Q/I$ and 
$U/I$) applying dual channel demodulation \citep{ishikawa2014b}, with plate scales of 0.0048~nm pixel$^{-1}$ and 
1.1~arcsec pixel$^{-1}$. Additionally, since the rotating half-waveplate
was located in front of the slit and the fold mirror of the slit-jaw system behaved as a partial
linear polarizer, the CLASP slit-jaw system itself also provided images of the broadband $Q/I$ fractional polarization, 
in addition to the intensity images mentioned above.

\section{Observational Results}

During its ballistic flight on 2015 September 3, CLASP observed first a quiet region at 
the solar disk center during 17:02:53--17:03:09UT in order to quantify the instrumental polarization, 
and we confirmed that the spurious polarization is one order of magnitude smaller than 0.1~\% 
\citep{giono2016a,giono2017}. Afterward, during 17:03:36--17:08:25UT, CLASP observed relatively quiet regions of the 
Sun (Figure~1(Aa)), with the slit oriented radially from 20~arcsec off-limb toward the disk center.

Figure~1(B) shows the wavelength variation of Stokes $I$, $Q/I$ and $U/I$ measured by CLASP at 
each pixel along the slit. The intensity emission profile shows significant spatial fluctuations on 
scales $\sim$10~arcsec, which essentially corresponds to the chromospheric network that results from 
intense photospheric magnetic field concentrations and related heating events: bright in the network 
and dark in the internetwork \citep[Figure~1(Aa); see also][]{kubo2016}. In the wings, $Q/I$ is negative (the 
linear polarization is perpendicular to the limb) with amplitudes larger than 1~\% and increasing 
up to $\sim\,6$~\% toward the limb (i.e., with a very clear center-to-limb variation, hereafter CLV). 
The $U/I$ wing signals are also very significant ($>1$~\%), with positive and negative values fluctuating 
around zero on spatial scales of $\sim\,$10~arcsec; they are a clear observational signature of 
the breaking of the axial symmetry (around the solar local vertical) 
of the Lyman-$\alpha$ radiation in the solar chromosphere, 
caused by the horizontal atmospheric inhomogeneities of the solar chromospheric plasma.
At the line center, the $Q/I$ and $U/I$ signals on 
the solar disk are $\sim$0.1~\%, and fluctuating around zero on spatial scales of $\sim\,$10~arcsec. 
However, at the off-limb positions sampled by the spectrograph's slit 
the $Q/I$ line-center signals are predominantly positive 
(parallel to the solar limb) with amplitudes of about $+0.5$~\%. Note that although weak ($\sim$ 0.1~\%), 
the measured $Q/I$ and $U/I$ line-core signals are an order of magnitude larger than the 
instrumental polarization level \citep{ishikawa2014b,giono2016a,giono2017} of CLASP. 

The overall shape of the $Q/I$ profiles (Figure~2), as well as the very significant negative 
$Q/I$ signals in the wings and their clear CLV (Figures~1(B) and 3), are essentially 
consistent with the theoretical predictions \citep{belluzzi2012}. The $Q/I$ and $U/I$ line-center amplitudes 
($<1$~\%) also follow the theoretical expectations for moderate (${\sim}3$ arcsec) 
spatial resolution observations \citep{trujillobueno2011,stepan2015}, such as those obtained by CLASP. However, 
the $Q/I$ line-center signals do not show any clear CLV, in marked contrast with the results of radiative transfer calculations 
in one-dimensional (1D) and three-dimensional (3D) models of the solar atmosphere \citep{trujillobueno2011,stepan2015}. 

Spatial fluctuations on scales of $\sim 10$~arcsec are seen both in intensity ($I$) and linear polarization 
($Q/I$, $U/I$). Maxima and minima of intensity along the slit in the line core are systematically 
shifted from those in the wings (Figure~3), the shift becoming more pronounced closer to 
the limb (especially at slit locations above 100~arcsec). This effect is due to the different optical 
depths in the core and wings, and hence to the different geometrical heights they sample (higher 
in the core, deeper in the wings). Interestingly, maxima of intensity often match minima of total 
linear polarization (${\rm LP}=\sqrt{Q^2+U^2}/I$) on the Lyman-$\alpha$ wings and vice versa, as shown by 
the colored segments in Figure~4(A). The anticorrelation between $I$ and LP is even more noticeable 
after removing from LP its CLV, proportional to $(1-\mu^2)$. In some places, even in the Lyman-$\alpha$ 
line core, LP is anti-correlated with the intensity (Figure~4(C)). The fact that regions with lower 
line intensity are, on average, more polarized than the high-intensity regions had been predicted 
by radiative transfer investigations of the Lyman-$\alpha$ scattering polarization in 3D
models of the solar atmosphere \citep{stepan2015}.

The CLASP slit-jaw system also, serendipitously, provided images of the broadband $Q/I$ 
fractional polarization (Figure~1(Ab)), in addition to the intensity images (Figure~1(Aa)). Since the 
Lyman-$\alpha$ $Q/I$ amplitudes are much larger in the wings than in the core, the broadband $Q/I$ signal that results 
from the wavelength-integrated Stokes $I$ and $Q$ profiles is dominated by the polarization 
in the wings. Interestingly enough, a comparison of this $Q/I$ image with the intensity image 
clearly shows that the low-intensity regions (e.g., quiet-Sun areas) are indeed more polarized 
than the high-intensity regions (e.g., plage areas).

\section{Discussion}

In general, there is a good agreement between the theoretical predictions and the spectropolarimetric 
observations obtained by CLASP. This is far from trivial since the quantum theory of 
polarization in spectral lines taking into account partial frequency redistribution phenomena is still under development.
Many of the subtle spectroscopic and radiative transfer effects it predicts (and explains) take place only in the 
extremely rarefied but generally optically thick plasma of a stellar atmosphere 
\citep[e.g.,][]{landi1998,stenflo2000,mansosainz2003}, and cannot be investigated 
through specifically dedicated laboratory experiments.

The $Q/I$ pattern observed by CLASP, with its strong ($>1$~\%) near-wing 
polarization signals, demonstrates that the underlying physical mechanism is the joint action of 
partial frequency redistribution and $J$-state interference (i.e., in the scattering events the photon 
goes through the two upper $J$ levels of Lyman-$\alpha$ at the same time). Without $J$-state interference, the $Q/I$ wing 
signals would be significantly weaker \citep[see][]{belluzzi2012}, contrary to the observations by CLASP. 

The complex polarization patterns shown in Figure~1(B) and especially the very significant polarization 
amplitudes at the Lyman-$\alpha$ wings are very sensitive to the thermal structure of the upper solar 
chromosphere, while the line core polarization is sensitive (via the Hanle effect) to the presence 
of weak (10--100 Gauss) magnetic fields in the enigmatic TR. The lack of CLV in $Q/I$ at line center is, however, 
an important and intriguing surprise, whose theoretical interpretation must significantly alter our empirical understanding 
of the solar TR. This is because the significant CLV at the center of the Lyman-$\alpha$ $Q/I$ profile 
predicted by state-of-the-art 3D models of the solar atmosphere \citep[see][]{stepan2015} can be made to disappear 
either by increasing the magnetization and/or the geometrical complexity of the model's chromosphere--corona TR, 
which is intimately related to the magnetic-field topology (\v{S}t\v{e}p\'{a}n et al. 2017, in preparation; 
Trujillo Bueno et al. 2017, in preparation).  

Finally, the non-trivial fact that the upper chromosphere and TR 
of the Sun generate significant linear polarization signals in the (optically thick) 
hydrogen Lyman-$\alpha$ line of the solar disk radiation supports the theoretical expectation 
that other ultraviolet lines with complementary magnetic sensitivities, 
such as the He {\sc ii} line at 30.4 nm \citep{trujillobueno2012} and the Mg {\sc ii} h \& k lines around 280.0 nm 
\citep{belluzzitrujillo2012}, should also be linearly polarized by anisotropic radiation pumping processes 
and that vacuum-ultraviolet spectropolarimetry is a key gateway to the magnetism and geometry of the
outer atmosphere of the Sun and other stars. To fully open up 
this new diagnostic window, we need high-precision instruments like CLASP on board space telescopes.

\acknowledgments

The CLASP team is an international partnership between NASA Marshall Space Flight Center, 
National Astronomical Observatory of Japan (NAOJ), Japan Aerospace Exploration Agency 
(JAXA), Instituto de Astrof\'{i}sica de Canarias (IAC), and Institut d'Astrophysique Spatiale; 
additional partners are the Astronomical Institute ASCR, Istituto Ricerche Solari Locarno (IRSOL), 
Lockheed Martin, and University of Oslo. The U.S. participation was funded by NASA Low Cost Access to Space (award 
number 12-SHP 12/2-0283). The Japanese participation was funded by the basic research 
program of ISAS/JAXA, internal research funding of NAOJ, and JSPS KAKENHI grant 
numbers 23340052, 24740134, 24340040, and 25220703. The Spanish participation was 
funded by the Ministry of Economy and Competitiveness through project AYA2010-18029 
(Solar Magnetism and Astrophysical Spectropolarimetry). The French hardware participation 
was funded by Centre National d'Etudes Spatiales (CNES). J.T.B. and J.S. gratefully acknowledge 
the supercomputing grants provided by the Barcelona Supercomputing Center (National Supercomputing Center, Barcelona, Spain).

\clearpage

\clearpage
\epsscale{1.0}

\begin{figure}
\includegraphics[angle=90,width=\hsize]{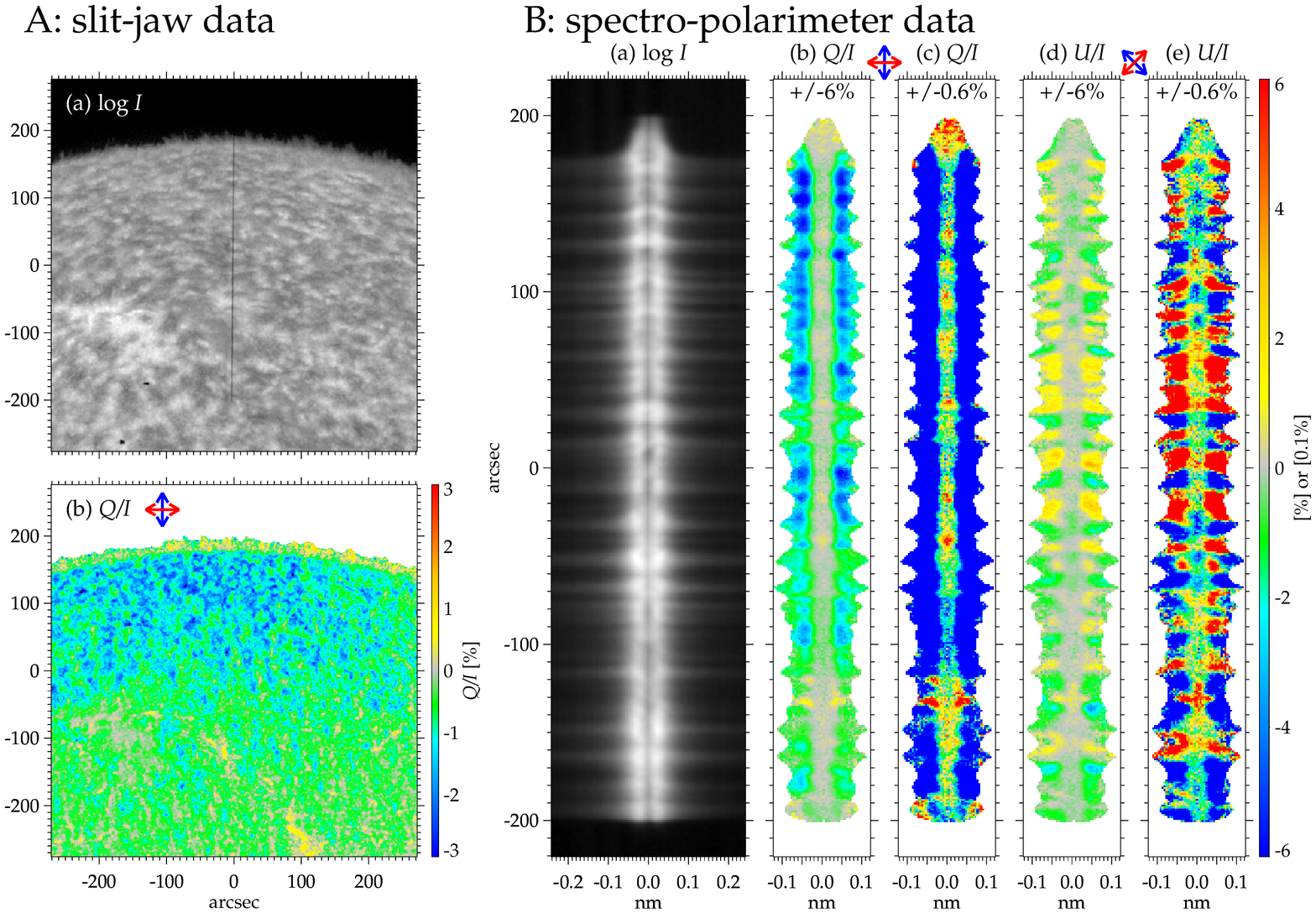}
\caption{CLASP data. (A) Broadband Lyman-$\alpha$ (a) intensity and (b) $Q/I$ images taken by the CLASP slit-jaw 
camera. The black line in (a) indicates the spectrograph's slit, which covers 400~arcsec. 
(B) Variation along the slit of the intensity ($I$) and 
fractional polarization ($Q/I$ and $U/I$) profiles of the hydrogen Lyman-$\alpha$ line 
observed by the CLASP spectropolarimeter. The solar limb is at $+175$~arcsec. (a) Stokes $I$ in logarithmic 
scale, (b, c) Stokes $Q/I$, and (d, e) Stokes $U/I$. The fractional polarization is clipped between 
$\pm 6$~\% in (b) and (d) for an optimal visualization of the wings and between $\pm 0.6$~\% in (c) and 
(e) for the line core. The reference directions for positive Stokes $Q$ and $U$ are indicated by the red arrows 
in the corresponding (B) panels.}
 \label{fig:1}
\end{figure}

\clearpage

\begin{figure}
\includegraphics[angle=90,width=\hsize]{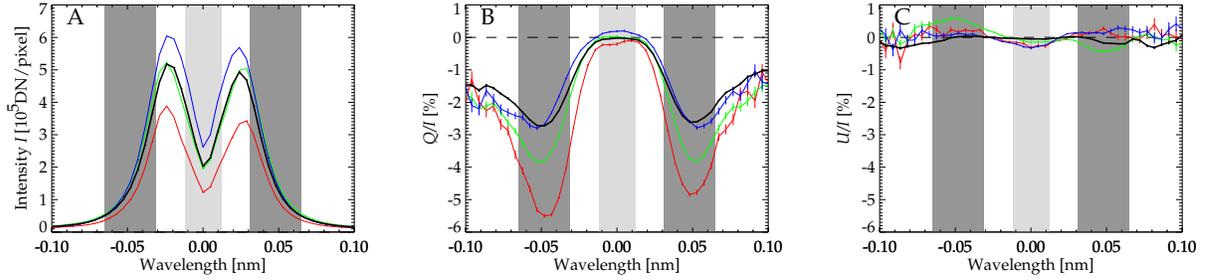}
 \caption{Examples of $I$, $Q/I$, and $U/I$ profiles observed by CLASP. (A) Intensity, (B) Stokes $Q/I$, and (C) Stokes $U/I$ 
at three different positions on the solar disk: $+152$'' (red), $+86$'' (green), and $-36$'' (blue) in 
Figure~1(B); these correspond to $\mu\equiv\cos\theta=0.2$, $0.4$, and $0.6$, respectively ($\theta$ being the heliocentric 
angle). To increase the signal-to-noise ratio, 21 pixels (23~arcsec) along the slit have been averaged. The 
vertical error bars show the standard deviation. The spatial average between 
$-198$'' and $+164$'' in Figure~1(B) is shown by the black curves.}
 \label{fig:2}
\end{figure}

\clearpage

\begin{figure}
\includegraphics[width=0.9\hsize]{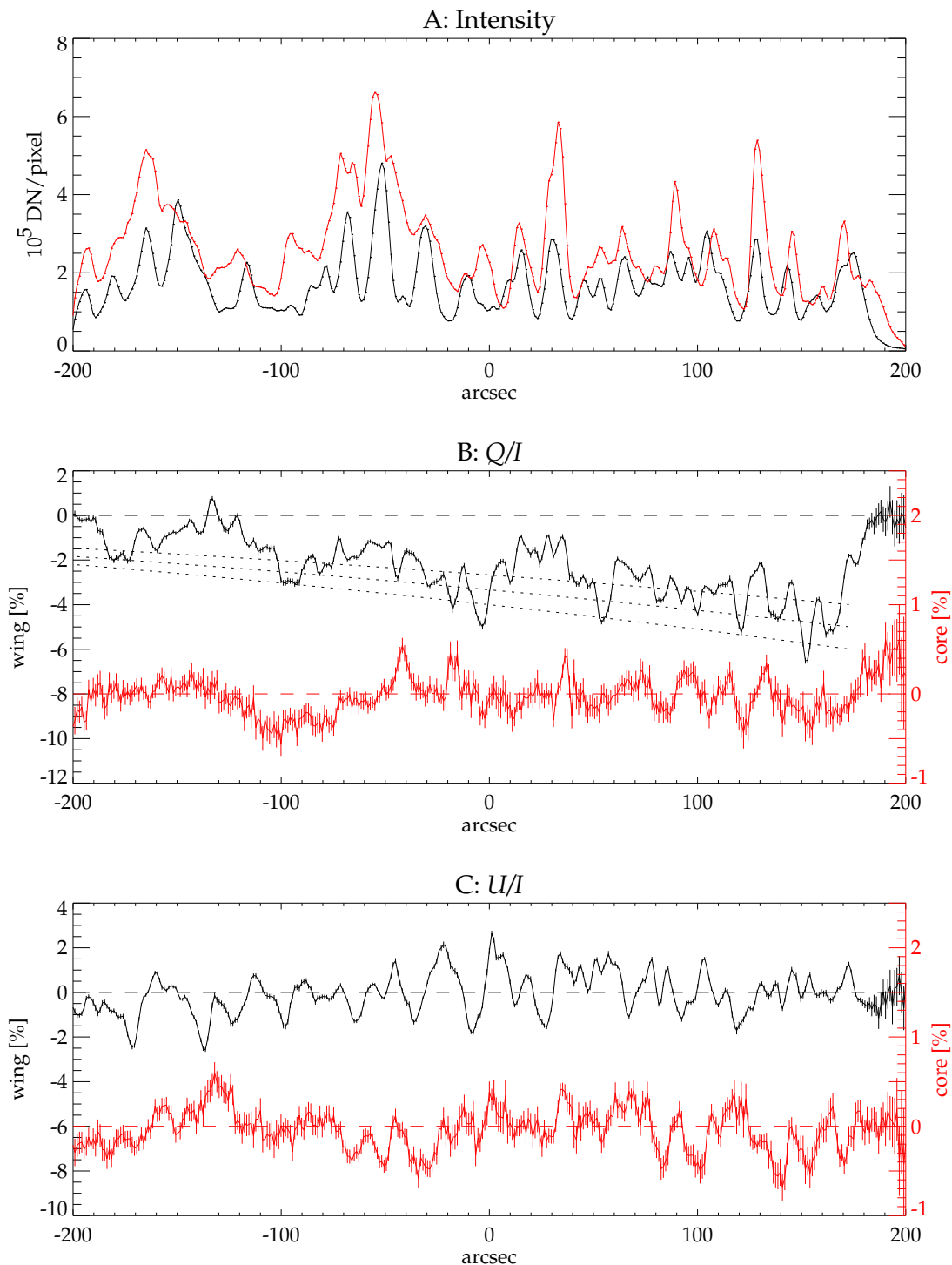}
 \caption{Spatial variations along the slit. (A) Intensity, (B) Stokes $Q/I$ and (C) Stokes $U/I$. The black 
curves correspond to the Lyman-$\alpha$ wing intervals given by the dark gray areas of Figure~2, while 
the red curves correspond to the Lyman-$\alpha$ core interval given by the light gray area of Figure~2. 
The dotted lines in panel (B) indicate the $(1-\mu^2)$ CLV trend.}
 \label{fig:3}
\end{figure}

\clearpage

\begin{figure}
\includegraphics[angle=90,width=\hsize]{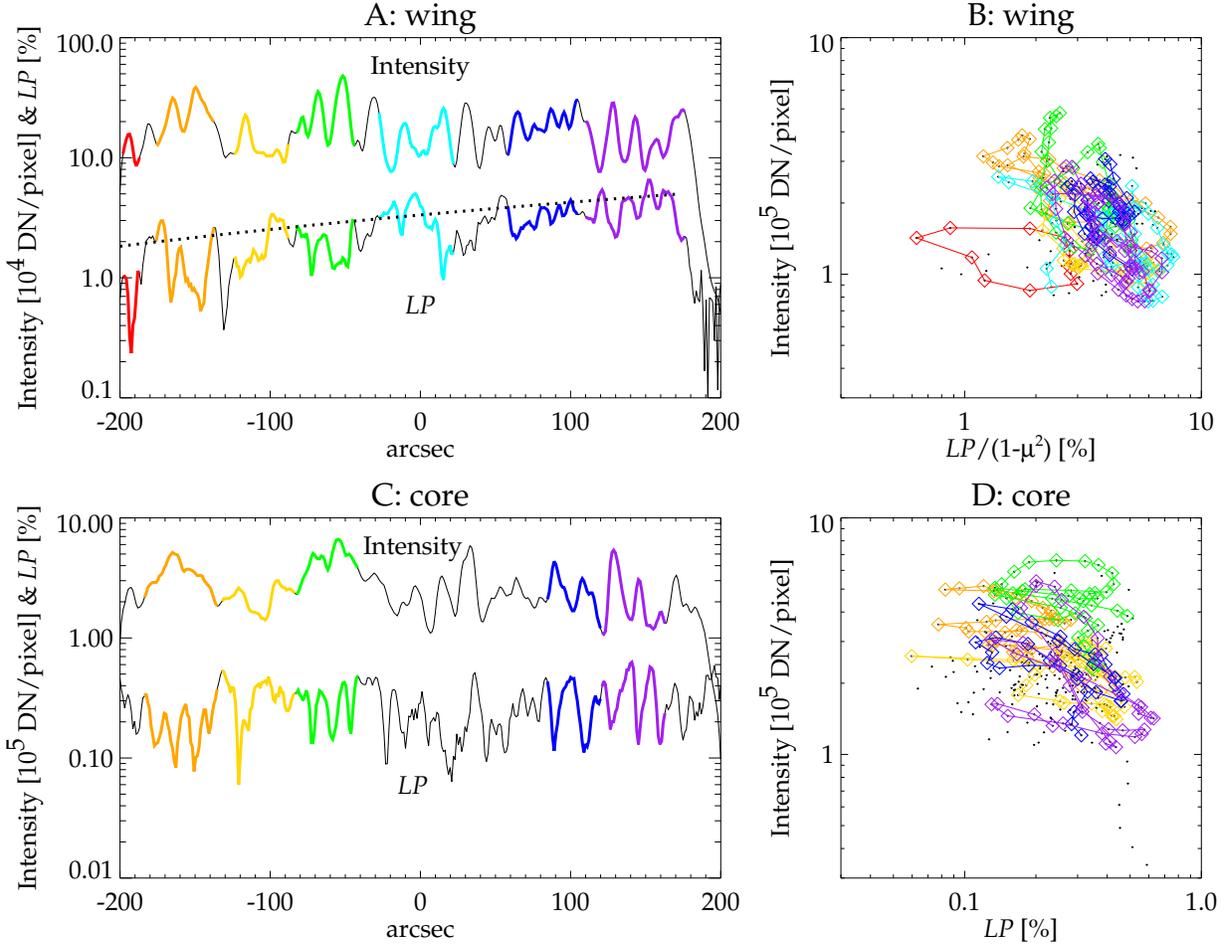}
 \caption{Spatial correlation between the intensity and the linear polarization degree LP. 
(A) The spatial variation of the intensity (upper curve) and 
LP (lower curve) at the Lyman-$\alpha$ wing. The segments having a clear anticorrelation 
between them are indicated by colors. (B) Scatter plot between LP and the intensity at the 
Lyman-$\alpha$ wing. The same colors are used for the data points in the colored segments. (C) and 
(D) are the same for the Lyman-$\alpha$ core. Only in panel (B) LP is divided by $(1-\mu^2)$ to remove 
the CLV trend.}
 \label{fig:4}
\end{figure}


\end{document}